\documentclass[12pt]{article}
\usepackage{amstex,eepic,float,epsfig}
\newcommand{\q}[4]{{n}_{#1#2#3}^{#4}}

\begin{document}
\title{Faddeev-Hopf knots: Dynamics of linked un-knots} \author{Jarmo
Hietarinta \\ Department of Physics, University of Turku\\ FIN-20014
Turku, Finland\\and\\ Petri Salo\\Helsinki Institute of Physics,
P.O. Box 9 (Siltavuorenpenger 20 C)\\ University of Helsinki,
FIN-00014 Helsinki, Finland} 
\maketitle

{\abstract We have studied numerically Faddeev-Hopf knots, which are
defined as those unit-vector fields in $R^3$ that have a nontrivial
Hopf charge and minimize Faddeev's Lagrangian.  A given initial
configuration was allowed to relax into a (local) minimum using the
first order dissipative dynamics corresponding to the steepest descent
method. A linked combination of two un-knots was seen to relax into
different minimum energy configurations depending on their charges and
their relative handedness and direction.  In order to visualize the
results we plot certain gauge-invariant iso-surfaces.}

\vskip 0.5cm The theory of knots has a long history in mathematics and
physics and research on this topic has again become very active,
partially due to the possible real world applications (DNA knotting
etc).  In the late 19th century knots were proposed as models of
atoms, and although this idea did motivate the mathematical study of
knots, it did not last as a model for physical reality. Now, with the
advent of string theory, the possibility of knotted elementary
structures has been considered again even in particle physics. In
order to give knots a physical meaning one has to give them some
physical properties, e.g., thickness and energy. For example, one may
assume that there is a charge distribution on the wire and then find
its minimum energy configuration~\cite{Si}.

The starting point for the knots discussed in this paper is classical
field theory with a unit-vector field defined everywhere in the
3-dimensional space.  The unit vector can be considered as a point in
$S^2$; furthermore, if at spatial infinity all vectors point to the
same direction (lets say to the north pole, i.e., ${\bf n}=(0,0,1)$)
we can compactify $R^3$ to $S^3$ and the vector field provides a map $
{\bf n}: S^3 \to S^2.  $ Such maps can be labeled by their Hopf
charge, $\pi_3(S^2)={\Bbb Z}$.  As an example we mention the following
map with Hopf charge one:
\begin{eqnarray}
n^1&=&\frac{4(2xz-y(x^2+y^2+z^2-1))}{(1+x^2+y^2+z^2)^2},\nonumber\\
n^2&=&\frac{4(2yz+x(x^2+y^2+z^2-1))}{(1+x^2+y^2+z^2)^2},\label{X:T}\\
n^3&=&1-\frac{8(x^2+y^2)}{(1+x^2+y^2+z^2)^2}.\nonumber
\end{eqnarray}
(It is easy to see that ${\bf n}\to (0,0,1)$ as $x^2+y^2+
z^2\to\infty$, as required). A knot can be associated to the vector
field by considering how it turns around. In particular, one can look
at the pre-image of the south pole ${\bf n}=(0,0,-1)$; normally it is
a closed curve which may form a knot.  For the vector field
(\ref{X:T}) the pre-image of the south pole is the ring
$z=0,\,x^2+y^2=1$, justifying the name ``un-knot''. Later it will
become clear that the curve forming the (un-)knot has also direction
and handedness, and the relevant knot theory is therefore that of
``framed links''~\cite{Fc}. [An un-knot of charge $Q$ is obtained from
(\ref{X:T}) if we replace $x\to\Re[(x+iy)^Q]/r^{Q-1},
y\to\Im[(x+iy)^Q]/r^{Q-1}$, where $r=\sqrt{x^2+y^2}$.]

In addition to topology we need physics, in particular we need to
define an energy for the field configuration. One requirement is that
it should not be feasible to minimize energy by scaling the knotted
configuration to zero or infinite size.  One early proposal was
\cite{Ni} $E={\tfrac12}\int [(\partial_\mu {\bf n})^2]^{\frac32} d^3x$
($\mu$-sum over 1 to 3), but this does not contain a scale. In
\cite{Fa} Faddeev proposed the Lagrangian
\begin{equation} 
L={\tfrac12}\int\left\{(\partial_\mu {\bf n})^2 + g
F_{\mu\nu}^2\right\} d^3x, \quad F_{\mu\nu}= \epsilon_{abc}
n^a\partial_\mu n^b \partial_\nu n^c,
\label{FL}
\end{equation}
as the energy functional. (The topological term can also be written as
$ F_{\mu\nu}^2=\frac12(\partial_\mu {\bf n} \times \partial_\nu {\bf
n})^2 =\frac12 [(\partial_\mu {\bf n})^2(\partial_\nu {\bf n})^2-
(\partial_\mu {\bf n} \cdot \partial_\nu {\bf n})^2]$.) What is
important here is that there is now a scale defined by the dimensional
coupling constant $g$.  Indeed, the derivatives mean that the
integrated kinetic term scales under $r\to\lambda r$ as $\lambda$, and
the $F^2$ term as $\lambda^{-1}$.  Thus a given initial knotted field
configuration (e.g. (\ref{X:T})) should evolve (with suitable added
dynamics) to a stable minimized configuration of a given size and
shape. Vakulenko and Kapitanskii \cite{VK} obtained the lower limit
(VK bound)
\begin{equation}
E\ge c\,\sqrt{g} |Q|^{\frac34},
\label{VKb}
\end{equation}
for the energy given by (\ref{FL}), here $c$ is some constant, $g$ the
coupling constant in (\ref{FL}), and $Q$ the Hopf charge.

The first numerical results with Lagrangian (\ref{FL}) were obtained
by Faddeev and Niemi \cite{FN1,FN2} (for popularizations see
\cite{pop}) and Gladikowski and Hellmund \cite{GH}. They studied
mainly charge 1 and 2 un-knots, although the possibility of real knots
was briefly discussed in \cite{FN1,FN2}.  The purpose of this letter is to
describe the dynamics of {\em linked} un-knots. We find that under
minimization the linked un-knots can deform to other configurations
following the rules applicable to framed links (Kirby moves).

The results of \cite{FN1,FN2} were obtained by a computational method
that is different from the one used here. Furthermore, in those papers
a rotationally symmetric ansatz was used, because it reduces the
computations to the two-dimensional space containing the $z$
axis. However, this reduction also means that stability under
rotation-dependent perturbations cannot be probed.  It turns out that
such effects can be seen already for the charge 3 un-knot for which
the minimum energy configuration is not rotationally symmetric
\cite{BS}. Our method is well suited for knots of more complicated
topology and we will report some fairly complicated minimization
dynamics for linked un-knots.  In \cite{FN1,FN2} another
simplification was made in handling the unit-vector field. After all,
the three components of a unit vector are not independent and one can
instead use the two components defined by $w\equiv U+iV
=(n^1+in^2)/(1+n^3)$.  Unfortunately there is a numerical problem in
using $w$ instead of $\bf n$, due to the fact that the range of $U$
and $V$ is from $-\infty$ to $+\infty$.  The singular values are
attained along a curve forming the knot, i.e., where ${\bf
n}=(0,0,-1)$. It is clear that in a relaxation process it is very
expensive to try to move infinities.  Indeed, in \cite{FN1,FN2} the
infinite inertia of the $w=\infty$ ring was compensated by changing
the coupling constants to best fit the virial theorem.  The behavior
near this infinity ring is also doubtful, because controlling overflow
situations may cause unexpected effects.

To avoid the problems mentioned above we decided from the very
beginning to put the system on a cubic lattice and use unit-vector
fields themselves.  Thus, the unknowns were $\q{i}jk\alpha$, where
$i,j,k$ give the location on the lattice and $\alpha$ the component,
$\sum_{\alpha=1}^3 (\q{i}jk\alpha)^2=1,\,\forall i,j,k$.  For the
kinetic energy term derivatives of ${\bf n}$ were discretized on
links, that is
\[
\int(\partial_\mu {\bf n}(x,y,z))^2 d^3r\to \sum_{i,j,k,\alpha}
\left[(\q{i+1}jk\alpha-\q{i}jk\alpha)^2 +
(\q{i}{j+1}k\alpha-\q{i}jk\alpha)^2
+(\q{i}j{k+1}\alpha-\q{i}jk\alpha)^2 \right]\delta.
\]
For the potential term we discretized $F_{\mu\nu}$ as defined in 
(\ref{FL}) on the $(\mu,\nu)$ plaquette. To make it completely 
symmetric we used in $F_{xy}$, say, 
\begin{eqnarray*}
n^\alpha &\to& \frac14 (\q{i}jk\alpha +
\q{i+1}jk\alpha+\q{i+1}{j+1}k\alpha+\q{i}{j+1}k\alpha),\\ \partial_x
n^\alpha &\to& \frac1{2\delta}[ (\q{i+1}jk\alpha-
\q{i}jk\alpha)+(\q{i+1}{j+1}k\alpha-\q{i}{j+1}k\alpha)],\\ \partial_y
n^\alpha &\to& \frac1{2\delta}[
(\q{i}{j+1}k\alpha-\q{i}jk\alpha)+(\q{i+1}{j+1}k\alpha-
\q{i+1}{j}k\alpha)].
\end{eqnarray*}

Minimization of the energy was done using the steepest descent method:
${\bf n}_{new}={\bf n}_{old}+\delta t\,\nabla L$, which corresponds to
the dissipative dynamics $\dot{\bf n}=\nabla L$. This was sometimes
accelerated by taking into account also the gradient of the previous
step. The expression for the gradient was calculated from the
discretized Lagrangian using the symbolic algebra program Reduce. We
did not use Lagrange multipliers, but rather renormalized ${\bf
n}_{new}$ to preserve unit length. On the lattice boundary the vector
was fixed to ${\bf n}=(0,0,1)$.  The fact that the system was put into
a box with fixed boundaries introduced some boundary pressure. For
this reason we cannot exactly attain the prediction of the virial
theorem stating that the topological and kinetic energies should be
equal.

We did first some calculations on a $50^3$ and $100^3$ lattices, but
it soon became clear that even for un-knots the discretization effects
could sometimes change the outcome: If the lattice is too small some
neighboring vectors could turn anti-parallel and break the
topology. The present computations were done on a $240^3$ lattice
divided into $4^3$ processors on Cray T3E parallel machine.  Each
round of iteration took about 2 seconds.  Typical runs took around
60-70.000 iterations, i.e., 30-40 hours (for charge 1 and 2 linked
un-knots).  Also, to make sure there were no topology-breakup effects
we followed the nearest-neighbor differences of the vectors, in our
computations $\max\left|{\bf n}\cdot{\bf n}_{nn}-1\right|$ was large
only at the very beginning and even then the differences never
exceeded $0.1$ and later were typically $0.02$.

It is, in general, hard to visualize vector-fields in three
dimensions.  Often we associate flow lines to vector fields by
imagining how a test particle would move if the vector field would be
its velocity field.  In the present case, however, this method is not
relevant. The point is that there is a global gauge invariance and
field lines change in an essential way under such global rotations.

As was said before, the unit vector field provides a map $ R^3\sim S^3
\to S^2$.  We have one fixed direction, that of the vector at infinity
(${\bf n}=(0,0,1)$), which defines the north pole of $S^2$. The unit
vector at any other point is defined by its {\em latitude} and {\em
longitude}.  In our illustrations we have chosen to display the
pre-image of some latitude circle, that is, the surface where the
vector field has a fixed angle with the vacuum direction, i.e., where
$n^3$ has a fixed value.

The longitude is described by colors on the iso-latitude surface. This
is suitable because colors are commonly put on a circle anyway: red
$\to$ yellow $\to$ green $\to$ blue $\to$ magenta $\to$ red\footnote{
Unfortunately the coloring algorithm we used did not completely
support the color-circle idea: For example, even if the values 0 and 1
were to stand for the same color, a triangle with corner values 0, 0
and 1 was painted with the color corresponding to the average
$1/3$.}. The actual position of the 0 meridian is not important,
because it is a gauge dependent quantity.  The figures were made using
the program "funcs"~\cite{JR} running on a Silicon Graphics O2.

\begin{figure}[p]
\epsfig{file=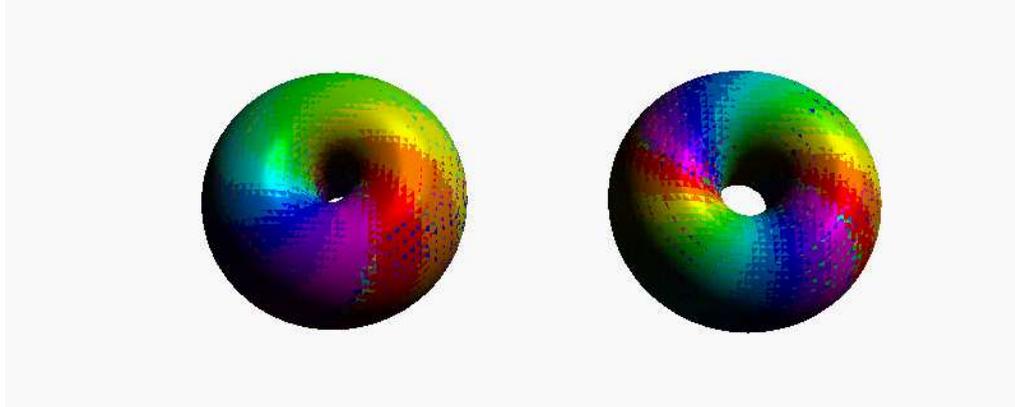}
\caption{ The final configurations for charge 1 and 2 un-knots. The
displayed iso-surfaces correspond to the equator, i.e., have $n^3=0$.}
\label{F:12}
\end{figure}

The vector field (\ref{X:T}) is not that suitable as an initial
configuration for numerical analysis because its kinetic energy
density decreases only as $r^{-4}$ as $r\to\infty$ and therefore the
integral itself as $r^{-1}$.  Thus, we introduced some radial
squeezing for large $r$ intended to minimize interaction with the box
boundaries, and some expansion for small $r$ to decrease the rapid
changes at the center. From such a starting configuration a single
un-knot relaxed fairly rapidly (in about 30-40.000 iterations) into
the minimum configuration. Usually we followed the minimization until
the changes in the energy were of the order $10^{-7}$ per iteration.
The same was done for the charge 2 un-knot. In both cases the final
configuration seemed to be rotationally symmetric, as can be seen from
Figure \ref{F:12}.  From these results we infer that for our
normalizations the constant $c$ in the VK bound (\ref{VKb}) is about
380.

In order to test stability under non-rotational perturbations we took
the final result of an un-knot (with charge 1 or 2), cut it into two
semi-tori, separated them by the distance of 20\% of the lattice
length by repeating the middle configuration, and embedded the new
configuration back into the original lattice. The relaxation quickly
took this configuration into the original circular form~\cite{Vid}.

From Figure \ref{F:12} one can clearly see that the un-knot has both
handedness and direction. The handedness can be determined e.g.\ by
holding the ring and checking with which hand the thumb is aligned
along the color boundaries. A right-handed ring (corresponding to
(\ref{X:T})) has positive Hopf charge.  We can arbitrarily associate a
direction to the positive charge un-knot, we use the color direction
red $\to$ yellow $\to$ green $\to$ blue $\to$ magenta $\to$ red. It
turns out that for the left-handed (=negative charge) un-knot we
should define the direction by the opposite color sequence.

Our main object was to study the relaxation dynamics of linked
un-knots.  The first problem was to construct suitable initial
configurations.  For a single un-knot we could use analytical formulae,
but for linked configurations this cannot be as easily done with one
formula. Besides, it would be better to use the already relaxed
numerical results as starting configurations. We therefore chose a
``cut and paste'' technique.

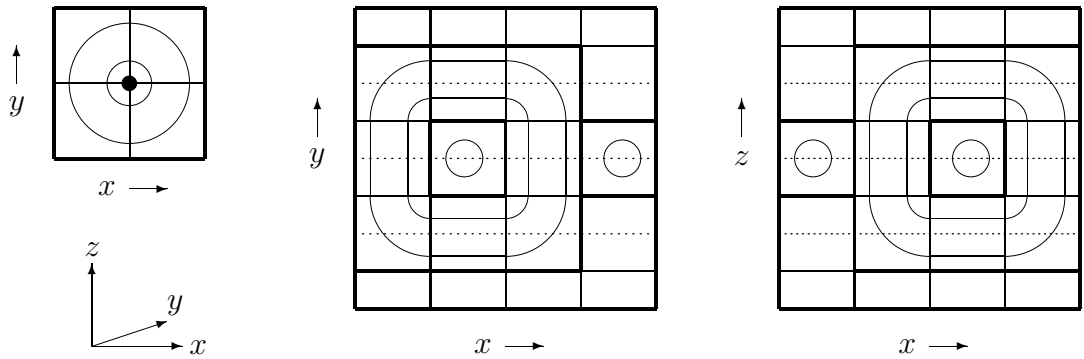
\begin{figure}[p]
\setlength{\unitlength}{1mm}
\begin{picture}(60,60)(-50,-10)
\linethickness{1pt}
\put(0,0){\line(1,0){40}}
\put(0,40){\line(1,0){40}}
\put(0,0){\line(0,1){40}}
\put(40,0){\line(0,1){40}}
\put(-40,20){\line(1,0){20}}
\put(-40,40){\line(1,0){20}}
\put(-40,20){\line(0,1){20}}
\put(-20,20){\line(0,1){20}}
\put(-30,30){\circle*{2}}
\linethickness{0.4pt}
\put(-40,30){\line(1,0){20}}
\put(-30,20){\line(0,1){20}}
\linethickness{0.2pt}
\put(-30,30){\circle{6}}
\put(-30,30){\circle{16}}
\put(-45,27){\makebox(0,0){$y$}}
\put(-45,30){\vector(0,1){5}}
\put(-33,16){\makebox(0,0){$x$}}
\put(-30,16){\vector(1,0){5}}
\put(-35,-5){\vector(1,0){12}}
\put(-35,-5){\vector(3,1){10}}
\put(-35,-5){\vector(0,1){11}}
\put(-22,-6){\makebox{$x$}}
\put(-25,0){\makebox{$y$}}
\put(-36,7){\makebox{$z$}}
\put(0,10){\dottedline{1}(0,0)(40,0)}
\put(0,20){\dottedline{1}(0,0)(40,0)}
\put(0,30){\dottedline{1}(0,0)(40,0)}
\linethickness{0.4pt}
\put(0,05){\line(1,0){40}}
\put(0,15){\line(1,0){40}}
\put(0,25){\line(1,0){40}}
\put(0,35){\line(1,0){40}}
\linethickness{0.4pt}
\put(10,0){\line(0,1){40}}
\put(20,0){\line(0,1){40}}
\put(30,0){\line(0,1){40}}
\linethickness{0.2pt}
\put(10,25){\arc{16}{3.145}{4.717}}
\put(10,25){\arc{6}{3.145}{4.717}}
\put(10,33){\line(1,0){10}}
\put(10,28){\line(1,0){10}}
\put(20,25){\arc{16}{4.717}{6.290}}
\put(20,25){\arc{6}{4.717}{6.290}}
\put(23,25){\line(0,-1){10}}
\put(28,25){\line(0,-1){10}}
\put(20,15){\arc{16}{0}{1.572}}
\put(20,15){\arc{6}{0}{1.572}}
\put(10,7){\line(1,0){10}}
\put(10,12){\line(1,0){10}}
\put(10,15){\arc{16}{1.572}{3.145}}
\put(10,15){\arc{6}{1.572}{3.145}}
\put(7,25){\line(0,-1){10}}
\put(2,25){\line(0,-1){10}}
\put(14.5,20){\circle{5}}
\put(35.5,20){\circle{5}}
\put(-5,20){\makebox(0,0){$y$}}
\put(-5,23){\vector(0,1){5}}
\put(17,-5){\makebox(0,0){$x$}}
\put(20,-5){\vector(1,0){5}}
\linethickness{1pt}
\put(10,15){\line(0,1){10}}
\put(20,15){\line(0,1){10}}
\put(10,15){\line(1,0){10}}
\put(10,25){\line(1,0){10}}
\put(0,5){\line(1,0){30}}
\put(0,35){\line(1,0){30}}
\put(30,5){\line(0,1){30}}
\put(30,15){\line(1,0){10}}
\put(30,25){\line(1,0){10}}

\end{picture}
\begin{picture}(60,60)(-45,-10)
\linethickness{1pt}
\put(0,0){\line(1,0){40}}
\put(0,40){\line(1,0){40}}
\linethickness{0.2pt}
\put(0,10){\dottedline{0.8}(0,0)(40,0)}
\put(0,20){\dottedline{0.8}(0,0)(40,0)}
\put(0,30){\dottedline{0.8}(0,0)(40,0)}
\linethickness{0.4pt}
\put(0,05){\line(1,0){40}}
\put(0,15){\line(1,0){40}}
\put(0,25){\line(1,0){40}}
\put(0,35){\line(1,0){40}}
\linethickness{1pt}
\put(0,0){\line(0,1){40}}
\put(40,0){\line(0,1){40}}
\linethickness{0.4pt}
\put(10,0){\line(0,1){40}}
\put(20,0){\line(0,1){40}}
\put(30,0){\line(0,1){40}}
\linethickness{0.2pt}
\put(20,25){\arc{16}{3.145}{4.717}}
\put(20,25){\arc{6}{3.145}{4.717}}
\put(20,33){\line(1,0){10}}
\put(20,28){\line(1,0){10}}
\put(30,25){\arc{16}{4.717}{6.290}}
\put(30,25){\arc{6}{4.717}{6.290}}
\put(33,25){\line(0,-1){10}}
\put(38,25){\line(0,-1){10}}
\put(30,15){\arc{16}{0}{1.572}}
\put(30,15){\arc{6}{0}{1.572}}
\put(20,7){\line(1,0){10}}
\put(20,12){\line(1,0){10}}
\put(20,15){\arc{16}{1.572}{3.145}}
\put(20,15){\arc{6}{1.572}{3.145}}
\put(17,25){\line(0,-1){10}}
\put(12,25){\line(0,-1){10}}
\put(4.5,20){\circle{5}}
\put(25.5,20){\circle{5}}
\put(-5,20){\makebox(0,0){$z$}}
\put(-5,23){\vector(0,1){5}}
\put(17,-5){\makebox(0,0){$x$}}
\put(20,-5){\vector(1,0){5}}
\linethickness{1pt}
\put(20,15){\line(0,1){10}}
\put(30,15){\line(0,1){10}}
\put(20,15){\line(1,0){10}}
\put(20,25){\line(1,0){10}}
\put(10,5){\line(1,0){30}}
\put(10,35){\line(1,0){30}}
\put(10,5){\line(0,1){30}}
\put(0,15){\line(1,0){10}}
\put(0,25){\line(1,0){10}}
\end{picture}
\caption{Constructing a linked combination from separate un-knots by
cut and paste. The vector field has the vacuum direction at the
boundaries marked by thick lines.}
\label{F:CP}
\end{figure}

Two un-knots each in a box of size $L^3$ (scaled, if necessary, to the
same $g$ value) are immersed into a box of size $(2L)^3$ as described
in Figure \ref{F:CP}. Assume that the ring of the first un-knot is on
the $(x,y)$-plane and at infinity the vectors point to the positive
$z$-direction. From this single un-knot configuration ${\bf n}$ we use
the part $0<x<L,\,0<y<L,\,\frac14L<z<\frac34L$. The new configuration
${\bf N}$ in the large cube contains, at first, eight pieces defined
by ${\bf n}$ as follows: (the variable range is always $
0<x<\frac12L,\,0<y<\frac12L, \frac14L<z<\frac34L$)
\begin{eqnarray*}
{\bf N}(x,y+\tfrac14L,z+\tfrac12L)&=&{\bf n}(x,y,z),\\
{\bf N}(x,y+\tfrac34L,z+\tfrac12L)&=&{\bf n}(x,\tfrac12L,z),\\ 
{\bf N}(x,y+\tfrac54L,z+\tfrac12L)&=&{\bf n}(x,y+\tfrac12L,z),\\
{\bf N}(x+\tfrac12L,y+\tfrac54L,z+\tfrac12L)&=
	&{\bf n}(\tfrac12L,y+\tfrac12L,z),\\
{\bf N}(x+L,y+\tfrac54L,z+\tfrac12L)&=
	&{\bf n}(x+\tfrac12L,y+\tfrac12L,z),\\
{\bf N}(x+L,y+\tfrac34L,z+\tfrac12L)&=
	&{\bf n}(x+\tfrac12L,\tfrac12L,z),\\
{\bf N}(x+L,y+\tfrac14L,z+\tfrac12L)&=&{\bf n}(x+\tfrac12L,y,z),\\
{\bf N}(x+\tfrac12L,y+\tfrac14L,z+\tfrac12L)&=&{\bf n}(\tfrac12L,y,z).
\end{eqnarray*}
The other un-knot is inserted similarly, but first it must be turned
so that it lies in the $(x,z)$-plane, by ${\bf n}'(x,y,z)={\bf
n}(x,z,L-y)$. (Note that this only moves the vectors without changing
their directions.) The remaining empty sub-blocks are filled with
vectors pointing up. The above construction makes sense only if there
are no discontinuities at the boundaries. In particular, it is
necessary that also at the center of the initial un-knot the vector
points up: ${\bf n} (\tfrac12L,\tfrac12L,z) =(0,0,1)$.

\begin{figure}[p]
\epsfig{file=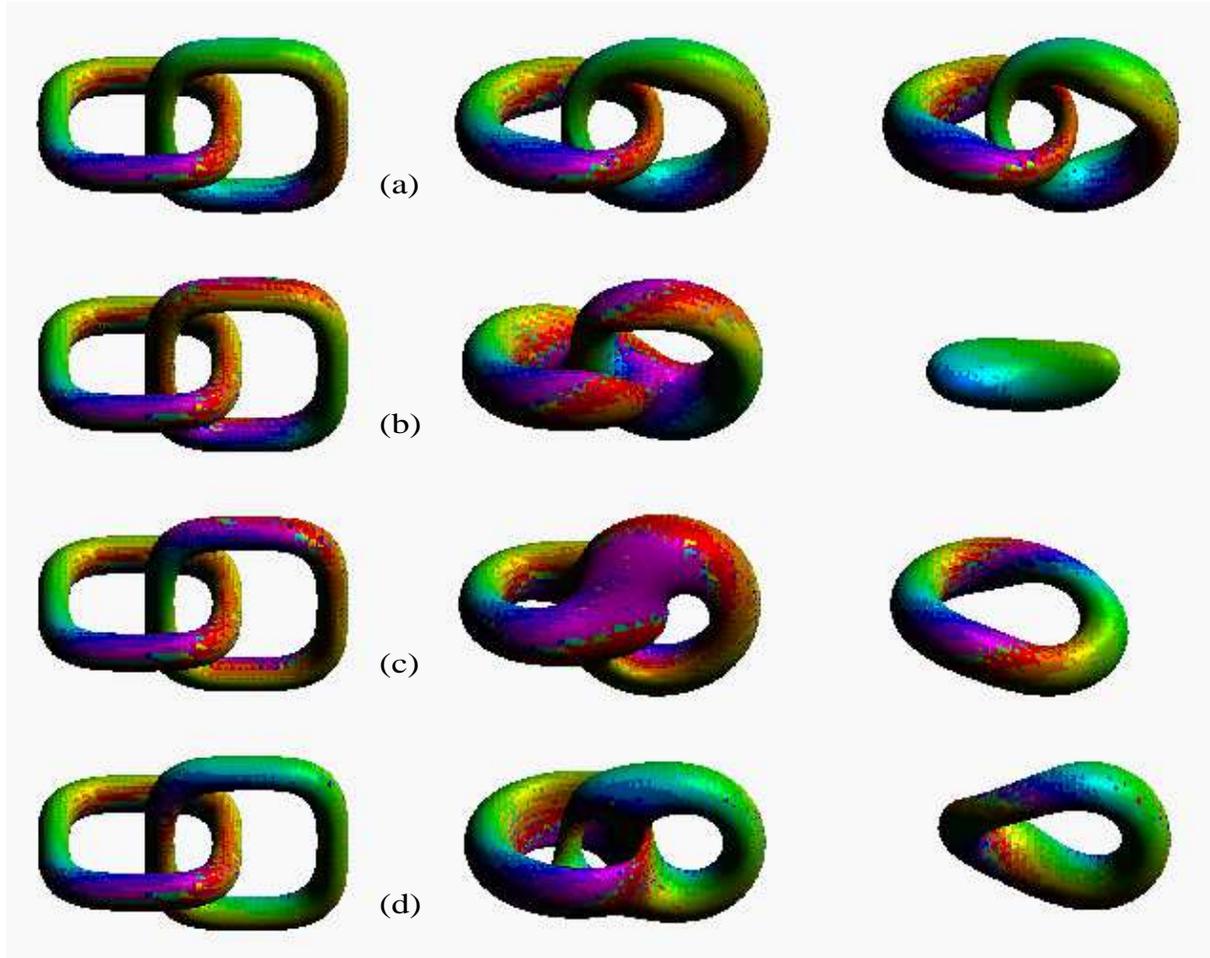,width=16cm}
\caption{Initial and final configurations for the four ways of
linking a charge $+1$ un-knot with a charge $\pm1$ un-knot. On the
left-hand column we have the initial configuration, at the center some
intermediate state, and on the right the ``final'' state (iso-latitude
surfaces $n^3=-0.5$). The charges and linking numbers are (a) $1+1+2$,
(b) $1+1-2$, (c) $1-1+2$, (d) $1-1-2$.}
\label{IFcol}
\end{figure}

In Figure \ref{IFcol} we show the initial and final results for some
linked configurations\cite{Vid}. On the left is the initial
configuration (constructed in the way described) after a few
iterations.  In each case the ring on the $(x,y)$-plane (on the left
side of the linked combination) is the same, it is right-handed with
Hopf charge $+1$ and has counterclockwise color direction if we look
from the positive $z$ direction. In the top two figures the part that
lies in the $(x,z)$-plane also has charge $+1$, but the parts are
linked differently, in (a) the linking is such that the color
direction is down through the un-knot in the $(x,y)$ plane and then the
linking number is $+2$, for (b) it goes up and the linking number is
$-2$, c.f., Figure \ref{Flink}. For the two lower figures the other
un-knot is left-handed (charge $-1$) and when we recall the different
definitions of direction mentioned above, the direction in (c) is from
above and for (d) from below.

The total charge of the linked configuration is the sum of the charges
of the initial un-knots and $\pm2$ for linking, and it is this total
charge that is conserved under minimization. For (b) the sum is zero
and indeed the final configuration is on its way to the vacuum where
all vectors point up. For (c) the sum is $+2$ and for (d) $-2$ which
are the charges of the final un-knots.  It should be noted that we did
not continue the minimization too long after the form of the final
configuration became clear, because it was already known that deformed
charge 2 un-knots return to their original shape.  For (a) the sum is
$+4$ and the configuration seems to be stable. The total energy of
this configuration is not too far from what is predicted by the VK
bound (within 10\%). A plot of energy vs.\ iteration step is given in
Figure \ref{Fene}.

The form of the final state depends on the relative color positions of
the two un-knots. This is illustrated in Figure \ref{Frot}.  The two
final configurations are different because the color orientations of
the initial un-knots in the $(x,z)$-plane were different,
corresponding to a $90^o$ rotation around the $y$ axis~\cite{Vid}. The
initial color difference has not changed much in the minimization,
instead we see that the un-knots are twisted differently.

We also studied the dynamics related to the four ways of combining
un-knots of charge $\pm2$~\cite{Vid}. The results agree with the addition
rule, their energy plots have been included in Figure
\ref{Fene}. Again the $2+2+2$ configuration seems to be stable, when
we stopped minimization its energy was within 7\% of the VK bound.

It should be emphasized that the minimization dynamics represents a
smooth deformation of the initial configuration. One way to explain
the observed deformations is to describe the configurations in terms
of framed links and then to deform them using Kirby moves\cite{Fc},
the details will be discussed elsewhere.

Recently Battye and Sutcliffe observed that higher charge un-knots
sometimes deform into more knotted but lower energy
configurations~\cite{BS}.  They found, e.g., that a perturbed charge 6
un-knot deformed into the linked combination $2+2+2$ discussed
above. For charge 4 they found a strongly deformed un-knot and it
would be interesting to check whether its final state energy really is
less than our $1+1+2$ configuration. Furthermore, they found that the
charge 7 un-knot deformed into a trefoil.  In order to test whether a
trefoil is also obtained form other initial configurations with total
charge 7 we constructed a linked combination of charge $+4$ and $+5$
un-knots with linking number $-2$. The initial state quickly deformed
into a wiggled charge 7 un-knot which then more slowly deformed into a
trefoil. On the other hand, the combination $2+3+2$ remained stable
and relaxed to equally low energies (within 5\%).

Our results show that the set of minimum energy configurations of
Faddeev-Hopf knots is very rich, and the minimization can take a very
complicated route. It would now be very interesting to characterize
further the topological properties of these types of knotted
configurations. In particular, it would be useful to know which
configurations of the same Hopf charge can be smoothly deformed to
each other, and whether in such cases there exists a deformation
sequence where energy always decreases. In any case we know now that
the associated knot theory must take into account the handedness and
direction of the curve forming the knot.

\section*{Acknowledgments}
We would like to thank L. Faddeev, A. Niemi, P. Sutcliffe and
W. Zakrzewski for discussions and M. Gr\"ohn for help in the
visualization aspects of this work. We are grateful to the Center for
Scientific Computing, Espoo, for computer time in the Cray T3E
parallel machine, where most of the computations reported here were
made.

\newpage

\parskip 0.3cm
\parindent 0pt

\section*{Figure captions}
{\bf Figure 1.} The final configurations for charge 1 and 2 un-knots. The
displayed iso-surfaces correspond to the equator, i.e., have $n^3=0$.

{\bf Figure 2.} Constructing a linked combination from separate
un-knots by cut and paste. The vector field has the vacuum direction
at the boundaries marked by thick lines.

{\bf Figure 3.} Initial and final configurations for the four ways of
linking a charge $+1$ un-knot with a charge $\pm1$ un-knot. On the
left-hand column we have the initial configuration, at the center some
intermediate state, and on the right the ``final'' state (iso-latitude
surfaces $n^3=-0.5$). The charges and linking numbers are (a) $1+1+2$,
(b) $1+1-2$, (c) $1-1+2$, (d) $1-1-2$.

{\bf Figure 4.} Rules for linking number from color direction.

{\bf Figure 5.} Energy as a function of the iteration step for various
linked combinations. A: $1+1+2$,  B: $1+1-2$,  C: $1-1+2$,  D: $1-1-2$, 
E: $2+2+2$,  F: $2+2-2$,  G: $2-2+2$,  H: $2-2-2$.

{\bf Figure 6.} Two final states for the $1+1+2$ configuration. The
initial states had different color orientation in the un-knot on the
$(x,z)$ plane (iso-latitude surfaces $n^3=-0.8$).

\newpage

\begin{figure}[p]
\setlength{\unitlength}{1mm}
\begin{picture}(60,30)(-50,-10)
\put(0,15){\vector(-1,0){3}}
\put(-4,5){\vector(1,0){3}}
\put(-1,12){\vector(0,-1){3}}
\put(21.5,8.5){\vector(0,1){3}}
\Thicklines
\put(0,10){\ellipse{20}{8}}
\put(10,10){\arc{20}{2.9}{8.9}}
\put(-10,-8){\makebox{Linking number $+2$.}}
\end{picture}
\begin{picture}(60,30)(-40,-10)
\put(0,15){\vector(-1,0){3}}
\put(-4,5){\vector(1,0){3}}
\put(-1,8.5){\vector(0,1){3}}
\put(21.5,12){\vector(0,-1){3}}
\Thicklines
\put(0,10){\ellipse{20}{8}}
\put(10,10){\arc{20}{2.9}{8.9}}
\put(-10,-8){\makebox{Linking number $-2$.}}
\end{picture}
\caption{Rules for linking number from color direction.}
\label{Flink}
\end{figure}

\begin{figure}[p]
\epsfig{file=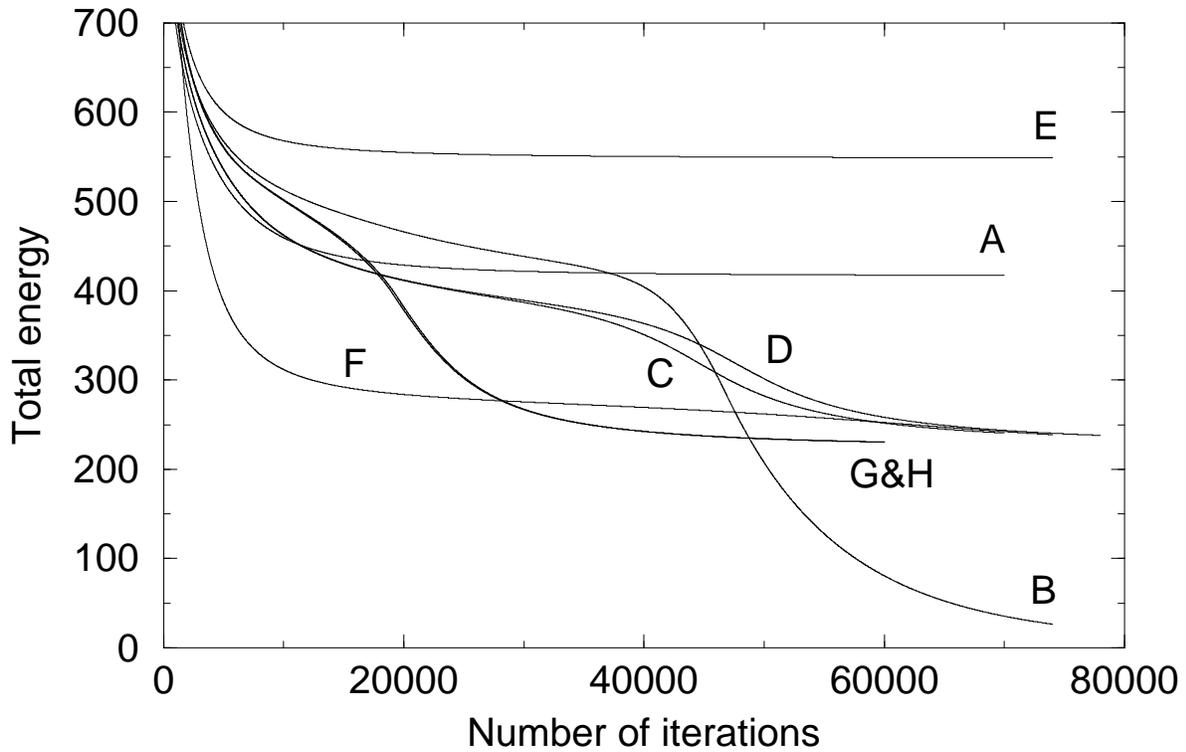,width=16cm}
\caption{Energy as a function of the iteration step for various
linked combinations. A: $1+1+2$,  B: $1+1-2$,  C: $1-1+2$,  D: $1-1-2$, 
E: $2+2+2$,  F: $2+2-2$,  G: $2-2+2$,  H: $2-2-2$.}
\label{Fene}
\end{figure}

\begin{figure}[p]
\epsfig{file=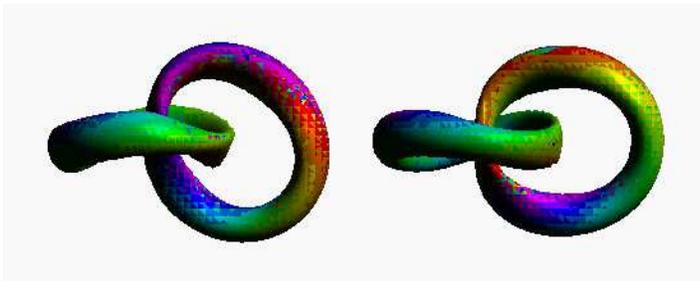}
\caption{Two final states for the $1+1+2$ configuration. The
initial states had different color orientation in the un-knot on the
$(x,z)$ plane (iso-latitude surfaces $n^3=-0.8$).}
\label{Frot}
\end{figure}

\end{document}